# Inelastic Neutron Scattering Studies of Phonon Spectra and Simulations in Tungstates, $AWO_4$ (A = Ba, Sr, Ca and Pb)

Prabhatasree Goel[1], M. K. Gupta[1], R. Mittal[1], S. Rols[2], S. N. Achary[3], A. K.Tyagi[3] and S. L. Chaplot[1]

[1]*Solid State Physics Division, Bhabha Atomic Research Centre, Mumbai 400085, India*
[2]*Institut Laue-Langevin, BP 156, 38042 Grenoble Cedex 9, France*
[3]*Chemistry Division, Bhabha Atomic Research Centre, Mumbai 400085, India*

Lattice dynamics and high pressure phase transitions in $AWO_4$ (A = Ba, Sr, Ca and Pb) have been investigated using inelastic neutron scattering experiments, *ab-initio* density functional theory calculations and extensive molecular dynamics simulations. The vibrational modes that are internal to $WO_4$ tetrahedra occur at the highest energies consistent with the relative stability of $WO_4$ tetrahedra. The neutron data and the *ab-initio* calculations are found to be in excellent agreement. The neutron and structural data are used to develop and validate an interatomic potential model. The model is used for classical molecular dynamics simulations to study their response to high pressure. We have calculated the enthalpies of the scheelite and fergusonite phases as a function of pressure, which confirms that the scheelite to fergusonite transition is second order in nature. With increase in pressure, there is a gradual change in the $AO_8$ polyhedra, while there is no apparent change in the $WO_4$ tetrahedra. We found that that all the four tungstates amorphize at high pressure. This is in good agreement with available experimental observations which show amorphization at around 45 GPa in $BaWO_4$ and 40 GPa in $CaWO_4$. On amorphization, there is an abrupt increase in the coordination of the W atom while the bisdisphenoids around A atom are considerably distorted. The pair correlation functions of the various atom pairs corroborate these observations. Our observations aid in predicting the pressure of amorphization in $SrWO_4$ and $PbWO_4$, which have not been experimentally reported.

I. INTRODUCTION

Compounds of the form MWO$_4$ (M = Mn, Cu, Ba, Sr *etc*) are studied[1-11] extensively for their interesting properties and usefulness. In this paper we report lattice vibrational studies on AWO$_4$ type tungstates with A = Ba, Sr, Ca and Pb, which are important for their scientific and technological applications[12-18]. These compounds in nano or bulk state find applications in solid- state scintillators[16-19], optoelectronic devices, solid state laser applications etc as well as in understanding of geological aspects. They also form a part of oxide ceramic composites useful for cryogenic detectors. These possible applications have motivated extensive interest in understanding the fundamental physical properties[20-31] of the AWO$_4$ tungstates.

The compounds under normal conditions crystallize in the tetragonal scheelite[28] structure (CaWO$_4$ mineral structure). The scheelite structure (space group 88, Z = 4) adopted by AWO$_4$ has a body centered unit cell, with A atoms occupying $S_4$ symmetry while the sixteen oxygen atoms occupy $C_1$ sites. Each W is surrounded by four equivalent oxygen atoms in a tetrahedral symmetry and A is surrounded by eight oxygen atoms. Along a-axis the WO$_4$ units are directly aligned whereas along the c-axis, AO$_8$ dodecahedra are interspersed between two WO$_4$ tetrahedra. Hence, different arrangement of hard WO$_4$ tetrahedra along the **c** and **a**-axis accounts for the difference in compressibility of the two unit-cell axes. These compounds exhibit a rich phase diagram with respect to change in pressure and temperature[23]. Several experimental and theoretical efforts have been undertaken to obtain understanding of the phase diagram of the compounds[20-27], which still remains elusive. Extensive Raman[27,28,29] and infrared[30] scattering techniques have been used to study the zone centre phonon modes in several of these tungstates. Experimental studies using angle dispersive X-ray diffraction (ADXRD), x-ray absorption near edge structure measurements (XANES) have observed that upon compression at high pressures AWO$_4$ undergoes a scheelite to fergusonite phase transition at room temperature. This transformation is a displacive in nature[23]. The β angle of the high pressure monoclinic cell of the fergusonite phase is only a little different from 90º. Upon further compression the compounds change into denser monoclinic and orthorhombic phases, before amorphizing. In the meanwhile, experiments like Raman scattering suggest[27-29] that there is a possibility of a reconstructive, first order transition to P2$_1$/n phase which coexists with the fergusonite phase. But this competing phase is kinetically hindered.

Preliminary results of inelastic neutron scattering measurements on SrWO$_4$ and theoretical studies on SrWO$_4$ and BaWO$_4$ have been communicated by us in conference proceedings[32]. Herein we report complete and comparative details of the inelastic scattering experiments and theoretical studies on a series of tungstates as AWO$_4$ (A = Ba, Sr, Ca and Pb). The measurement of phonons density of states is carried out by using neutron inelastic scattering. The transferable interatomic potential has been developed and further used in molecular dynamics studies to understand the mechanism of pressure driven phase changes of AWO$_4$ (M = Ba, Ca, Sr and Pb). We have looked at the changes occurring in the lattice under increasing pressure. The simulations have enabled us to understand the behavior of various polyhedra under compression.

## II. EXPERIMENTAL DETAILS

Polycrystalline samples of the AWO$_4$ (A= Ba, Sr and Ca) were prepared by solid state reaction of appropriate amounts of alkaline earth carbonate (ACO$_3$) and WO$_3$. PbWO$_4$ was prepared by heating the precipitates obtained from Pb(NO$_3$)$_2$ and (NH$_4$)$_2$WO$_4$ solutions. The phonon density of states measurements of all AWO$_4$ (A= Ba, Sra, Ca, Pb) were performed in the neutron-energy gain mode with incident neutron energy of 14.2 meV. The data were taken over the scattering angle 10$^\circ$ to 113$^\circ$. The signal is corrected for the contributions from the empty cell and suitably averaged over the angular range using the available software package at ILL. The incoherent approximation[33] was used in the data analysis. The data were suitably averaged over the angular range of scattering using the available software package at ILL to obtain the neutron-cross-section weighted phonon density of states. The multiphonon contribution has been calculated using the Sjolander[34] formalism and subtracted from the experimental data.

## III. THEORETICAL FORMALISM

*Ab-initio* calculations were performed using the projector-augmented wave (PAW) formalism of the Kohn-Sham density functional theory (DFT) at the generalized gradient approximation level (GGA) formulated by the Perdew-Burke-Ernzerhof (PBE) density functional[35-37]. We have used Vienna *Ab-initio* Simulation Package (VASP) for calculation of total energy and force[38,39]. The force constants have been calculated using super cell approach implemented in PHONON package[40], which are further used for calculation of phonon frequencies in the entire Brillioun zone. The total energy calculation has been done on 14 different configurations generated by displacement of symmetry inequivalent atoms along different Cartesian directions ($\pm x, \pm y, \pm z$). The kinetic energy cutoff for total energy calculations is 860 eV for all the four compounds.

We have used a transferable interatomic potential to study the vibrational properties of the tungstates using lattice dynamics calculations. An interatomic potential[41-43] consisting of short range and Coulombic terms is used. The radii parameters for M atoms are $r_{Ca}$ = 2.06 Å, $r_{Ba}$ = 2.4 Å, $r_{Pb}$ = 2.25 Å and $r_{Sr}$ = 2 Å respectively. The parameters of the WO$_4$ stretching potential are D = 3.2 eV, $r_o$ = 1.75 Å and n = 26.2 Å$^{-1}$. The van der Waals interaction between O-O pairs is 80 eVÅ$^6$. The polarizibility of the oxygen and alkali earth atoms has been considered in the framework of shell model[43] with shell charge and shell core force constants for oxygen atoms as −2.0 and 50 eV/Å$^2$ and for alkaline atoms as 3.0 and 50 eV/Å$^2$ respectively. Molecular dynamics (MD) simulations have been carried out using the model parameters obtained from the lattice dynamics calculations. A super cell consisting of 8×8×4 unit cells (6144 atoms) has been used to study the response of the compounds on compression. Rigid ion model has been used in the MD simulations. Calculations have been carried out from ambient pressure to around 50 GPa at room temperature. Equation of state has been integrated for about 1400 picoseconds using a timestep of 0.002 picoseconds. All the calculations have been done using the lattice and molecular dynamics softwares[44] DISPR and MOLDY developed at Trombay, Mumbai.

## IV. RESULTS AND DISCUSSION

The neutron inelastic scattering experiments have been performed using powder samples of the AWO$_4$ (M = Ba, Sr, Ca and Pb) at ambient conditions. The measurements have been carried out on the time of flight (TOF) instrument IN4C, ILL, Grenoble. The measured phonon density of states of the four tungstates in comparison with ab-initio and potential model calculations are given in Fig. 1. The potential model calculations results (Fig. 1) are in good agreement with the inelastic neutron data and *ab-initio* calculations. This has allowed validation of our model, which is further used to explore the phase stability of the compounds using molecular dynamics simulations as discussed below. The measured spectra show that phonons occur in two distinct regions in these compounds. The first region extends from 0 to 60 meV, while there is another region between 90-120 meV. The higher energy component corresponds to the internal vibrations of the WO$_4$ tetrahedra.

The ab-initio calculated partial phonon densities of each atom of AWO$_4$ compounds are given in Fig. 2. It can be seen that the contribution from the A atoms extends up to 20 meV. The contribution of Pb atom is at the lowest energy while that of Ca at the highest energy. This is in commensurate with their relative atomic mass. The contribution from W atoms is confined within 20 meV, with a very small contribution up to 60 mev. There is another contribution above 90 meV in the band between 90-120 meV from W atoms. In case of oxygen, contribution is in the complete range 0-60 meV and in the band between 90 to 120 meV. The partial phonon densities ascertain that the contribution above 90 meV in the measured spectra is from the internal stretching modes of the W-O tetrahedra. The computed density of states has been used to determine the specific heat capacity of the four oxides as shown in Fig. 3. The available data on $CaWO_4$[45] has been compared and found to be in very good agreement with our calculated data.

Using the interatomic potential developed, we have studied the pressure evolution of the tungstates. Fig. 4 gives the equation of state of the four compounds. These compounds are found to undergo initial transformation from scheelite ($I4_1/a$) to fergusonite ($I2/a$). According to experimental XANES[23,24] and ADXRD[23,24] studies, it is now well known that $BaWO_4$ transforms from scheelite to fergusonite phase at 7.5 GPa, $PbWO_4$ transforms at 9 GPa, $CaWO_4$ undergoes the above said transition at 11 GPa, and $SrWO_4$ at 10 GPa. The pressure driven transition to fergusonite phase is of displacive in nature, it is known to be a second order phase transition with almost no discernable volume discontinuity. The change in the β angle is only feeble; change in the 'b' lattice parameters is also negligible. This change is not evident from our MD calculations.

In order to understand the nature of the scheelite to fergusonnite transtition, we have carried out enthalpy ($H = \Phi + PV$, where $\Phi$ is the crystal potential energy) calculations using ab-initio method in the two phases. The calculated difference in enthalpy between scheelite and fergusonite phase is shown in Fig. 5. We can see that difference in the enthalpy of the two phases increases gradually beyond a certain pressure value, which is different for each of the $AWO_4$. Our findings are in agreement with previous experimental XANES and ADXRD[23] measurements, that this transition is continuous and second order in nature.

In our molecular dynamics calculations we can see that there is a discontinuity in the pressure-volume curve (Fig. 4), which suggests a change in phase. In case of $CaWO_4$, this change is seen around 30 GPa, around 34 GPa in $SrWO_4$, in case of $PbWO_4$ it is around 38 GPa while in case of $BaWO_4$ it is around 45 GPa. Detailed studies were done to understand this sudden change in the equation of state. Fig. 6 gives the structural arrangement of the different polygons at different pressures (in $BaWO_4$) as obtained from our computations. At P = 0 GPa, the atoms are arranged in the scheelite structure, A atom is in $AO_8$ coordination while W is in tetrahedral $WO_4$ coordination. Regions B and C are identified to depict the coordination of A and W atoms. As pressure is increased, at 43 GPa, there is no visible change in the structural arrangement of the various polygons. There is no change in the coordination around A and W atoms. Beyond 45 GPa, there is a volume drop in $BaWO_4$. This observation coincides with the volume discontinuity observed around 45 GPa in $BaWO_4$ in Fig. 4. The structure at 47 GPa shows several obvious changes as compared to earlier structures. In region C, it can be seen that A atoms are no longer in perfect bisdisphenoids, instead it is now seen in a highly distorted polygon. In region B, W atoms show significant changes, the co-ordination number has increased. Most of the $WO_4$ tetrahedra have changed in to distorted $WO_6$.

Further credence to this observation is obtained from the O-W-O bond angle distribution given in Fig. 7. At ambient pressure, all $WO_4$ tetrahedra are regular with O-W-O angle equal to 109°. At 43 GPa, $WO_4$ tetrahedra are still regular. But on amorphization, the increase in coordination has resulted in the change in the bond angle. Majority of the O-W-O angle is around 90º which corresponds to an octahedral arrangement. Not all the W's have attained a coordination of 6; hence distortment of the tetrahedra has given rise to a range of angles for the O-W-O bond as seen in Fig. 7. Similar behavior is seen in all the remaining tungstates.

In order to obtain a deeper understanding into this volume discontinuity, we have studied the local order in the lattice between various atomic pairs. The pair correlations have been computed in these compounds with increasing pressure. Fig. 8 shows the pair-correlation function between A-W, A-O, A-A, W-W, W-O and O-O at various pressures. We find that with increasing pressure, there are subtle and small changes in the correlations between the different atomic pairs. In case of $BaWO_4$, the correlations broaden beyond 47 GPa, as seen from the plots in Fig. 8. This is in very good agreement with the reported value[24] of 45 GPa for amorphization in $BaWO_4$. In case of $CaWO_4$, the broadening of peaks appear around 32 GPa, while the experiments show[28] amorphization around 40 GPa. Our molecular dynamics calculations predict that in case of $SrWO_4$, at 36 GPa the peaks in the calculated pair correlations broaden, implying that the discrete ordering is lost. In $PbWO_4$, similar trend is observed at 40 GPa. These observations are in tandem with the equation of states (Fig. 4) computed in each of the compounds.

Careful investigations of the various atomic pairs reveal some interesting observations. In case of A-O pair distribution, we find that changes due to increase in pressure is seen to occur gradually, as can be seen in the inset of the Ba-O pair distribution. There is gradual decrease in the bond length on going from 0 GPa to 43 GPa. The $BaO_8$ units are able to withstand the increased pressure with gradual rearrangements of the Ba-O bonds, but beyond 43 GPa, the polyhedra distorts considerably. This leads to the broadening of peaks, suggesting the loss of long range order. In case of $WO_4$ tetrahedra, there is almost no change in the bond-length (changes by about 0.015 Å) as pressure is increased from 0 to 43 GPa (Inset Fig. 8). However further increase in pressure beyond a threshold of 45 GPa leads to an abrupt increase in the W-O bond by about 0.1 Å. Hence it can be inferred that $WO_4$ tetrahedra is almost immune to changes in pressure up to an upper threshold of pressure, whose value is different for each of the tungstate. Beyond this threshold, $WO_4$ tetrahedra distort and W atom's coordination increases.

Our above said observations are in tandem with experimentally[24] observed amorphization around 45 GPa in $BaWO_4$ and at around 40 GPa in $CaWO_4$[28]. The equation of state for various compounds has also been obtained (Fig. 4) from minimization of the enthalpy at 0 K. Table I gives the bulk modulus and its pressure derivative for the four compounds in comparison with the available reported experimental data[24]. The values are obtained from fitting of the Birch equation of state[46] to the calculated equation of states.

Earlier studies[23] surmise that large ratio of ionic radii ($WO_4$/A) accommodates increased stresses through larger and more varied displacement from their average positions. This results in the loss of translational periodicity at high-pressure. Larger the $WO_4$/A ratio, lower the pressure threshold for pressure induced amorphization. This ratio decreases as per the following order: $WO_4$/Ca>$WO_4$/Sr>$WO_4$/Pb>$WO_4$/Ba. As per this notion, $CaWO_4$ should have the lowest value of pressure for amorphization to occur, followed by $SrWO_4$, $PbWO_4$ and $BaWO_4$ respectively. Our computed results are in full agreement with this expected trend.

## V. CONCLUSIONS

A combination of model and *ab-initio* lattice dynamics has been used to study the vibrational properties in the four tungstates. Molecular dynamics simulations have been employed to understand their evolution with increasing pressure. Inelastic neutron scattering experiments have been carried out to measure the phonon density of states in the compounds. The calculations are in good agreement with experimentally measured data. With increasing pressure, tungstates undergo transition from scheelite to fergusonite phase, it is a second order displacive transition with no apparent volume discontinuity. Enthalpy calculations using ab-initio method, show that scheelite to fergusonite transition is second order in nature. But with further increase in pressure, these tungstates show pressure induced loss of translational ordering, which is seen by the sudden drop in volume in all the four tungstates studied, using molecular dynamics simulations. On amorphization $WO_4$ tetrahedra deform, the coordination of W atoms increases from 4. The $AO_8$ polyhedra show considerable distortion. The calculated amorphization pressures for $BaWO_4$ and $CaWO_4$ are in agreement with the available experimental data. The molecular dynamics simulations facilitates in predicting the amorphization in $SrWO_4$ and $PbWO_4$ at high pressures.

[1] D. Niermann, C. P. Grams, M. Schalenbach, P. Becker, L. Bohaty, J. Stein, M. Braden and J. Hemberger, Phys. Rev. B **89**, 134412 (2014).
[2] M. Baum, J. Leist, Th. Finger, K. Schmalzl, A. Hiess, L. P. Regnault, P. Becker, L. Bohaty, G. Eckold and M. Braden, Phys. Rev. B **89**, 144406 (2014).
[3] N. Poudel, K. C. Liang, Y. Q. Wang, Y. Y. Sun, B. Lorenz, F. Ye, J. A. Fernandez Baca and C. W. Chu, Phys. Rev. B **89**, 054414 (2014).
[4] I. V. Solovyev, Phys. Rev. B **87**, 144403 (2013).
[5] J. Ruiz Fuertes, S. Lopez Moreno, J. Lopez Solano, D. Errandonea, A. Segura, R. Lacomba Perales, A. Munoz, S. Radescu, P. Rodriguez-Hernandez, M. Gospodinov, L. L. Nagomaya and C. Y. Tu, Phys. Rev. B **86,** 125202 (2012).
[6] J. Ruiz-Fuertes, A. Segura, F. Rodriguez, D. Errandonea and M. N. Sanz Ortiz, Phys. Rev. Lett. **108**, 166402 (2012).
[7] S. Lopez-Moreno, P. Rodriguez-Hernandez, A. Munoz, A. H. Romero and D. Errandonea, Phys. Rev. B **84**, 064108 (2011).
[8] J. Ruiz Fuertes, D. Errnadonea, S. Lopez-Moreno, J. Gonzalez, O. Gomis, R. Vilaplana, F. J. Manjon, A. Munoz, P. Rodriguez Hernandez, A. Friedrich, I. A. Tupitsyna and L. L. Nagornaya, Phys. Rev. B **83**, 214112 (2011).
[9] F. J. Manjon, J. Lopez-Solano, S. Ray, O. Gomis, D. Santamaría-Pérez, M. Mollar, V. Panchal, D. Errandonea, P. Rodríguez-Hernandez, and A. Munoz, Phys. Rev. B **82**, 035212 (2010).
[10] J. Ruiz-Fuertes, D. Errandonea, R. Lacomba-Perales, A. Segura, J. González, F. Rodríguez, F. J. Manjon, S. Ray, P. Rodríguez-Hernández, A. Muñoz, Zh. Zhu, and C. Y. Tu, Phys. Rev. B **81**, 224115 (2010).
[11] R. Lacomba Perales, D. Errandonea, D. Martinez Garcia, P. Rodriguez Hernandez, S. Radescu, A. Mujica, A. Munoz, J. C. Chervin, and A. Polian, Phys. Rev. B **79**, 094105 (2009).
[12] R. Lacomba Perales, D. Martinez García, D. Errandonea, Y. Le Godec, J. Philippe, G. Le Marchand, J. C. Chervin, A. Polian, A. Munoz, and J. Lopez Solano, Phys. Rev. B **81**, 144117 (2010); V. Panchal, N. Garg, A. K. Chauhan, Sangeeta, and S. M. Sharma, Solid State Commun. **130**, 203 (2004).
[13] D. Errandonea, J. Pellicer Porres, F. J. Manjon, A. Segura, Ch. Ferrer Roca, R. S. Kumar, O. Tschauner, P. Rodríguez Hernández, J. Lopez Solano, S. Radescu, A. Mujica, A. Munoz, and G. Aquilanti, Phys. Rev. B **72**, 174106 (2005).
[14] Y. Zhang, N. A. W. Holzwarth, and R. T. Williams, Phys. Rev. B **57**, 12738 (1998).
[15] R. Lacomba-Perales, D. Errandonea, A. Segura, J. Ruiz-Fuertes, P. Rdriguez-Hernandez, S. Radescu, J. Lopez-Solano, A. Mujica and A. Munoz, J. Appl. Phys. **110**, 043703 (2011)
[16], A. A. Annenkov, M. V. Korzhik, and P. Lecoq, Nucl. Instrum. Methods, Phys. Res. A **490**, 30 (2002)
[17] M. Nikl, P. Bohacek, E. Mihokova, N. Solovieva, A. Vedda, M. Martini G. P. Pazzi, P. Fabeni, M. Kobayashi, and M. Ishii, J. Appl. Phys. **91**, 5041 (2002).
[18] A. Brenier, G. Jia, and C. Y. Tu, J. Phys.: Condens. Matter **16**, 9103 (2004).
[19] H. E. Chao, Y. Kuisheng, L. Lei and S. I. Zhenjun, J. Rare Earths **31,** 790 (2013).
[20] A Grzechnik, W A Crichton, M. Hanfled, Phys. Stat. Solidi **242** 2795 (2005).
[21] J. Lopez Solano, P. Rodriguez Hernandez, S. Radescu, A. Mujica, A. Munoz, D. Errandonea, F. J. Manjon, Pellicer Porres J, Garro N, Segura A, Ch. Ferrer Roca, R. S. Kumar, O. Tschauner and G. Aquilanti, Phys. Stat. Sol. B **244,** 325 (2007).
[22] F. J. Manjon, , D. Errandonea, J. Lopez Solano, P. Rodríguez Hernandez, S. Radescu, A. Mujica, A. Munoz, N. Garro, J. Pellicer Porres, A. Segura, Ch. Ferrer Roca, R. S. Kumar, O. Tschauner, and G. Aquilanti. Phys. Stat. Sol. (b**) 244,** 295 (2007).
[23] (a) F. J. Manjon and D. Errandonea, Prog. Mat. Sci. **53**, 711 (2008) and references therein. (b) P. Blanchfield and G. A. Saunders, J. Phys. C: Sol. Stat. Phys. **12**, 4673 (1979).

[24] D. Errandonea, J. Pellicer Porres, F. J. Manjon, A. Segura, Ch. Ferrer-Roca, R. S. Kumar, O. Tschauner, J. Lopez -Solano, P. Rodríguez-Hernandez, S. Radescu, A. Mujica, A. Munoz, and G. Aquilanti Phys. Rev. B **73,** 224103 (2006) and references therein.

[25] F. J. Manjon, D. Errandonea, N. Garro, J. Pellicer Porres, J. Lopez -Solano, P. Rodríguez-Hernandez, S. Radescu, A. Mujica and A. Munoz Phys. Rev. B **74,** 144111 (2006).

[26] F. J. Manjon, D. Errandonea, N. Garro, J. Pellicer-Porres, J. López-Solano, P. Rodríguez Hernandez, S. Radescu, A. Mujica, and A. Muñoz, Phys. Rev. B **74**, 144112 (2006).

[27] P. Rodriguez Hernandez, J. Lopez-Solano S. Radescu, A. Mujica, A. Munoz, D. Errandonea, F. J. Manjon, J. Pellicer Porres, A. Segura, Ch. Ferrer Roca, R. S. Kumar, O. Tschauner and G. Aquilanti, J. Phys. Chem. of solids **67**, 2164 (2006).

[28] D. Errandonea, M. Somayazulu and D. Hausermann, Phys. Stat. Sol. (b) **235**, 162 (2003)

[29] W. Sleight, Acta Crystallogr. B **28**, 2899 (1972).

[30] A. Jayaraman, B. Batlogg and L. G. VanViterrt, Phys. Rev. B **31**, 5423 (1985); **28**, 4774 (1983).

[31] P. J. Miller, R. K. Khanna and E. R. Lippincott J. Phys. Chem. Solids **34,** 533 (1973)

[32] Prabhatasree Goel, R. Mittal, S. L. Chaplot and A. K. Tyagi, Pramana, **71**, 1135 (2008); Prabhatasree Goel, R. Mittal and S. L. Chaplot, J. Phys.: conf. Series, **377**, 012094 (2012); Prabhatasree Goel et al, Chin. J. Phys., **49**, 308 (2011).

[33] D. L. Price and K. Skold, in *Methods of experimental physics: Neutron scattering Part A*, edited by K. Skold and D. L. Price (Academic Press, Orlando, 1986), Vol. **23**; J. M. Carpenter and D. L. Price, Phys. Rev. Lett. **54**, 441 (1985).

[34] A. Sjolander, Arkiv fur Fysik **14**, 315 (1958).

[35] S. Baroni, P. Giannozzi, and A. Testa, Phys. Rev. Lett. **58**, 1861 (1987).

[36] G. Kresse and J. Furthmüller, Comput. Mater. Sci. **6**, 15 (1996).

[37] G. Kresse and D. Joubert, Phys. Rev. B **59**, 1758 (1999).

[38] H. J. Monkhorst and J. D. Pack, Phys. Rev. B **13**, 5188 (1976)

[39] J. P. Perdew, K. Burke, and M. Ernzerhof, Phys. Rev. Lett. **77**, 3865 (1996).

[40] K. Parlinksi, Software phonon (2003).

[41] S. L. Chaplot, N. Choudhury, S. Ghose, M. N. Rao, R. Mittal and K. N. Prabhatasree, European Journal of Mineralogy **14**, 291 (2002).

[42] R. Mittal, S. L. Chaplot and N. Choudhury, Progress of Materials Science **51**, 211 (2006).

[43] G. Venkatraman, L. Feldkamp and V. C. Sahni, Dynamics of Perfect Crystals (MIT, Cambridge) (1975).

[44] S. L. Chaplot, Unpublished.

[45] W.G. Lyon and W.F. Edgar, J. Chem. Phys., **49** 3374 (1968).

[46] F. Birch, J. Geophys. Res. **57**, 227 (1952).

TABLE I : Comparison between the experimental and calculated bulk modulus (B) (in GPa units) and its pressure derivate (B'). The experimental values at 300 K are from Ref. [24]. The B values at 300 K has been obtained using the B' values estimated at 0 K.

|              | BaWO$_4$ |        | CaWO$_4$ |        | SrWO$_4$ |        | PbWO$_4$ |        |
|--------------|----------|--------|----------|--------|----------|--------|----------|--------|
|              | **B**    | **B'** | **B**    | **B'** | **B**    | **B'** | **B**    | **B'** |
| Expt. (300 K)| 52(5)    | 5(1)   | 74(7)    | 5.6(9) | 63(7)    | 5.2(9) | 66(5)    | 5.6(9) |
| Calc. (0 K)  | 55.8(8)  | 4.4(1) | 80.1(5)  | 4.5(1) | 69.9(2)  | 4.5(3) | 61.7(7)  | 5.1(1) |
| Calc. (300 K)| 46.8(4)  | 4.4    | 62.1(4)  | 4.5    | 56.3(6)  | 4.5    | 49.8(3)  | 5.1    |

FIG. 1 (Color Online) Neutron inelastic scattering data of the phonon density of states in AWO$_4$ (A = Ba, Ca, Sr and Pb) compared with our *ab-initio* and shell model calculations.

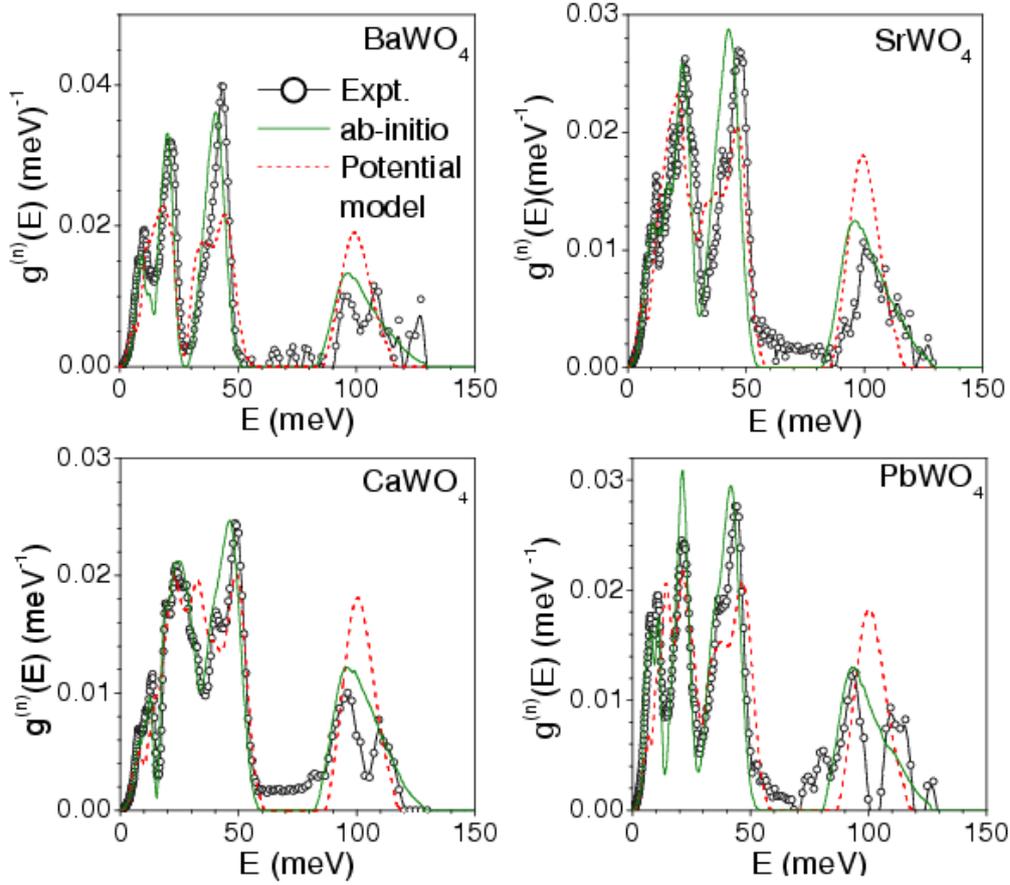

FIG. 2 (Color online) Ab-initio calculated partial densities of the constituent atoms in AWO$_4$ (A = Ba, Sr, Ca and Pb).

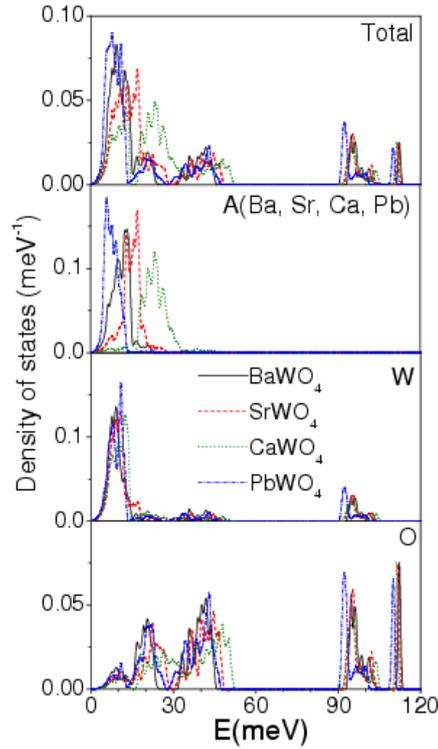

FIG. 3 Experimental and computed (using potential model) specific heat in AWO$_4$ (A = Ba, Sr, Ca and Pb). The experimental data for CaWO$_4$ is from Ref. [44].

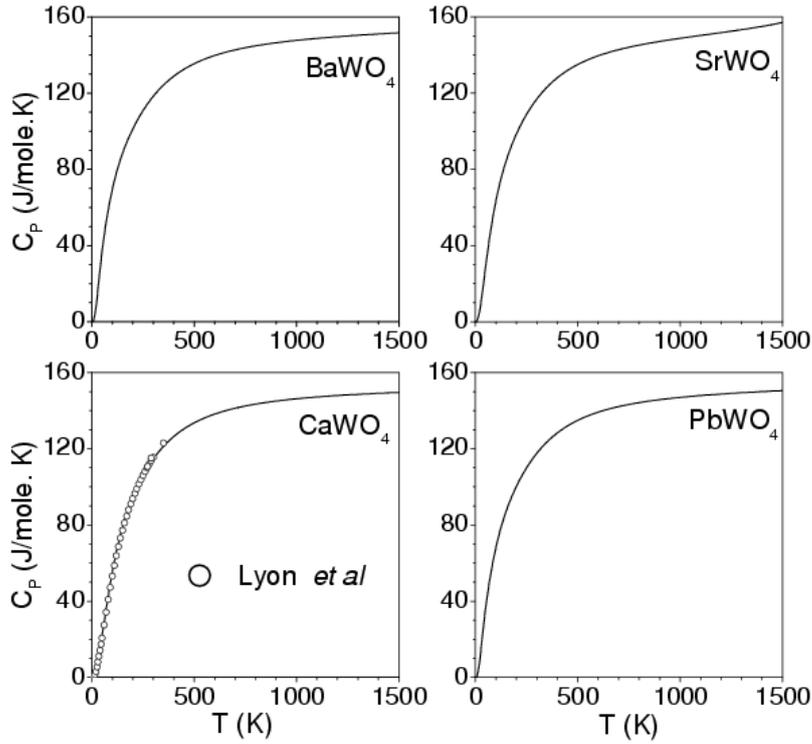

FIG. 4. Calculated (using potential model) and experimental equation of state of AWO$_4$ (A = Ba, Ca, Sr and Pb). V and V$_o$ are the unit cell volume at high and ambient pressure respectively. 'a', 'b' and 'c' correspond to the experimental data from references [13], [24] and [27] . The full and dash lines correspond to calculations at T=300 K and 0 K respectively.

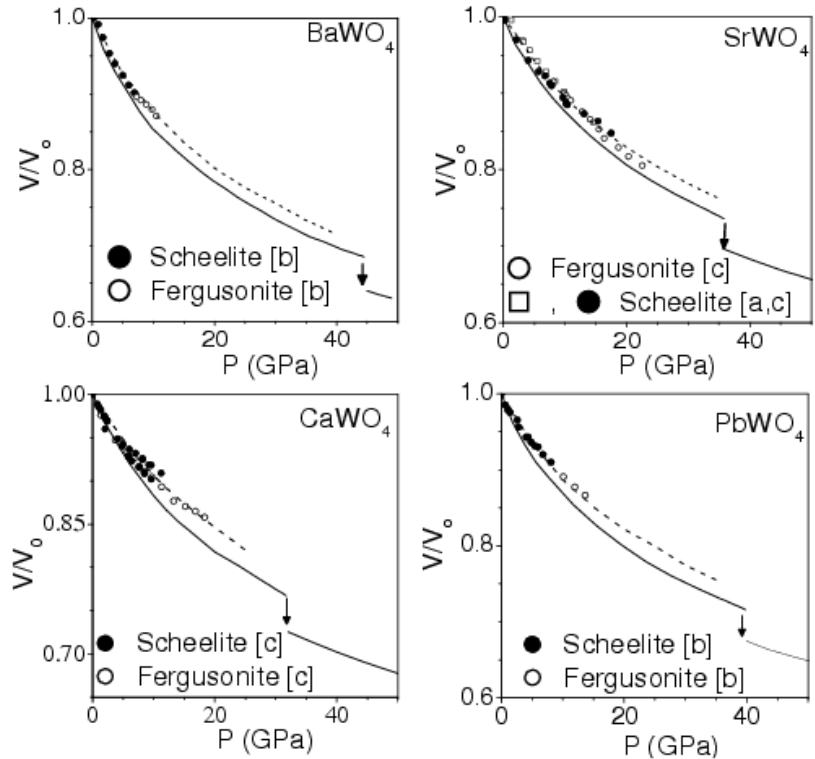

FIG. 5: Difference in enthalpy ($\Delta H = H_S - H_F$) with increasing pressure for the scheelite and fergusonite phases in all the four $AWO_4$ (A = Ba, Ca, Sr and Pb) compounds as computed from ab-initio studies.

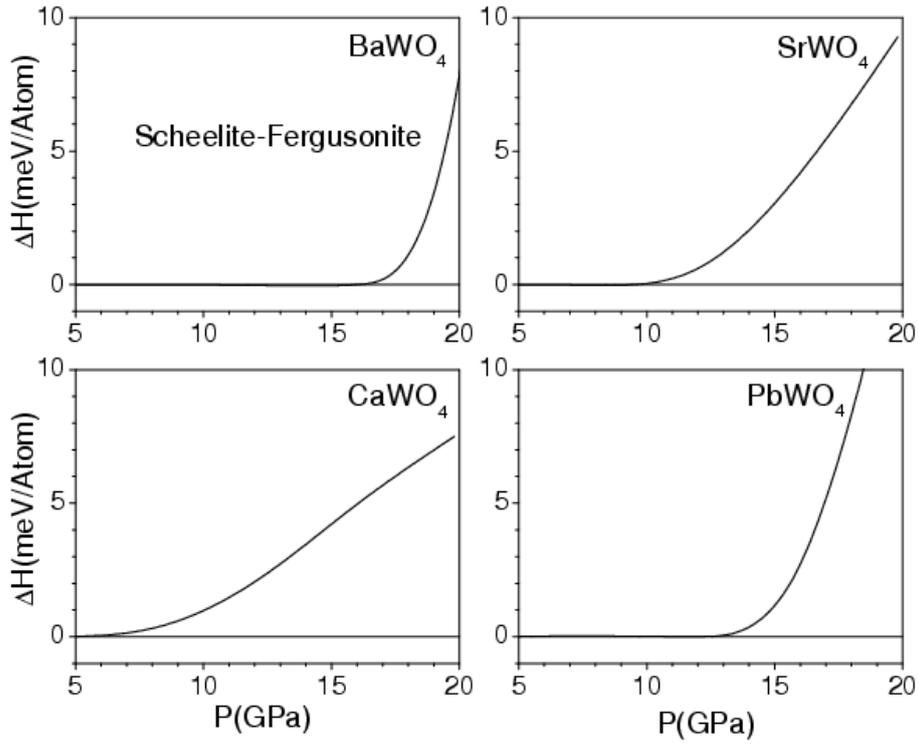

FIG. 6 (Color online): The structure of $BaWO_4$ super cell evolving under pressure. Regions B and C are marked to depict and understand the change in coordination around A and W atoms. (**c**-axis is perpendicular to the plane of the paper).

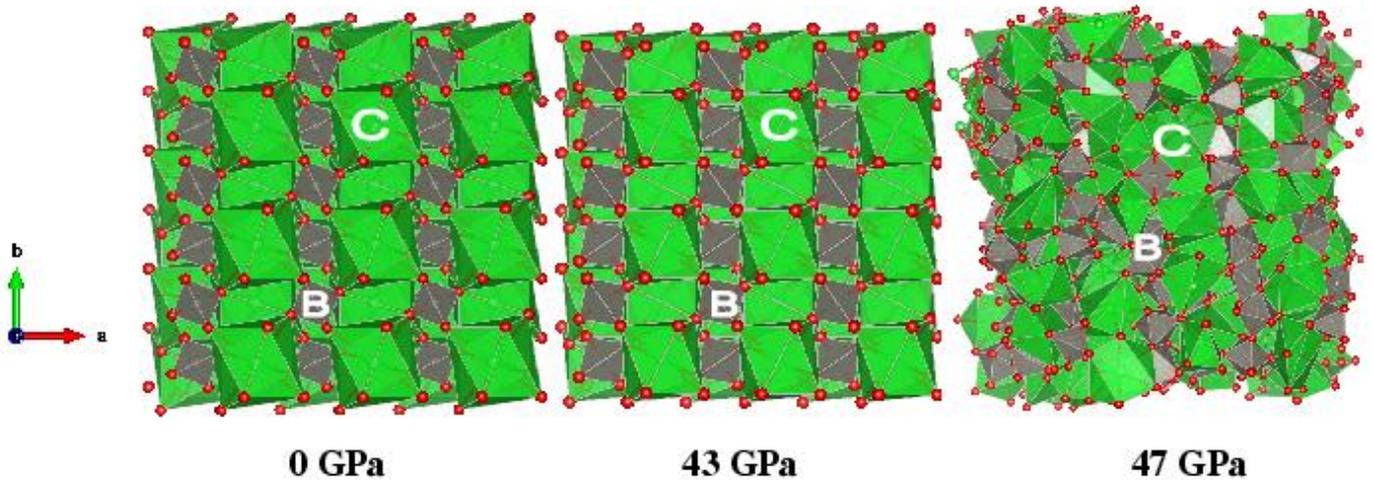

FIG. 7: Intratetrahedral bond angle O-W-O with changing pressure in BaWO$_4$.

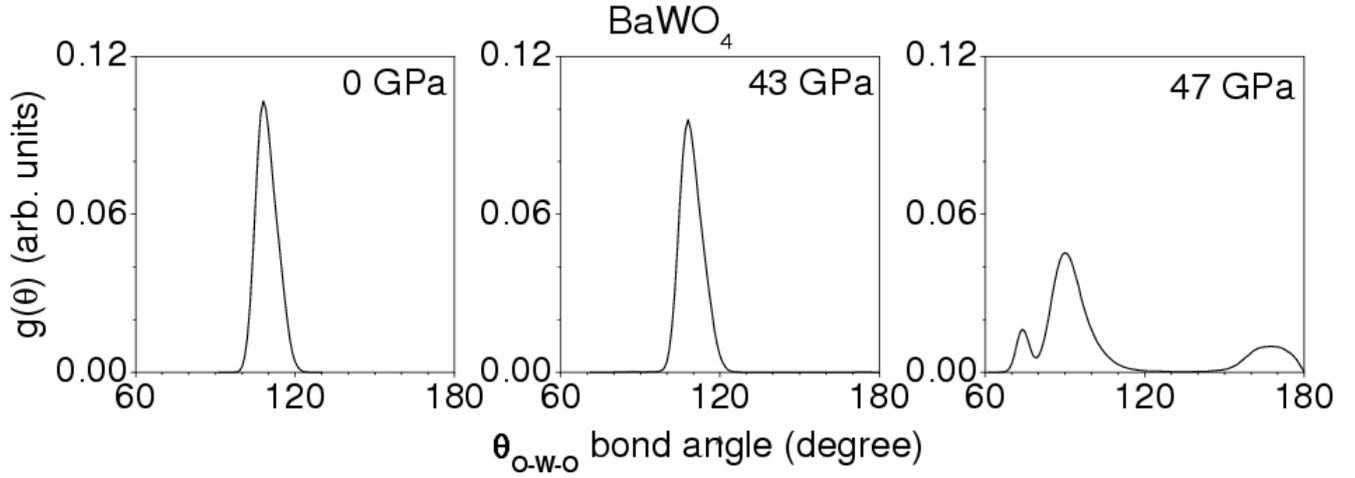

FIG. 8 (Color online) Pair correlations between various atomic pairs as obtained from molecular dynamics simulations.

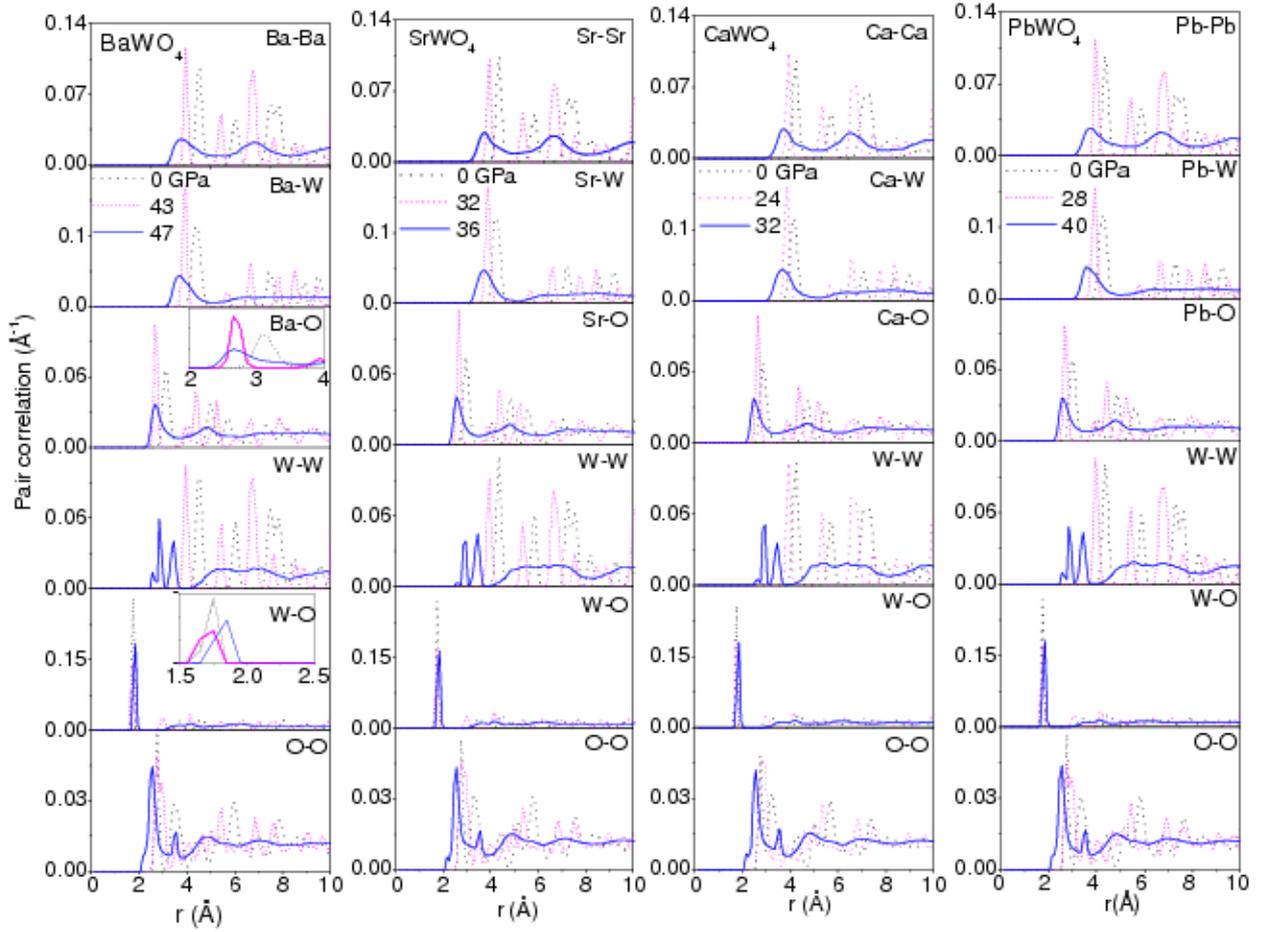